\documentclass[aps,prb,twocolumn,superscriptaddress,showpacs]{revtex4-1}

\usepackage{graphicx}
\usepackage{dcolumn}
\usepackage{bm}
\usepackage{hyperref}

\begin{document}


\title{Parallel Field Magnetoresistance in Topological Insulator Thin Films}



\author{C.~J.~Lin,$^1$ X.~Y.~He,$^1$ J.~Liao,$^1$ X.~X.~Wang,$^{1,2}$ V.~Sacksteder~IV,$^1$ W.~M.~Yang,$^1$ T. Guan,$^1$ Q.~M.~Zhang,$^1$ L.~Gu,$^1$ G.~Y.~Zhang}
\affiliation{Beijing National Laboratory for Condensed Matter Physics, Institute of Physics, Chinese Academy of Sciences, Beijing 100190, China}

\author{C.~G.~Zeng}
\affiliation{Hefei National Laboratory for Physics at Microscale and Department of Physics, University of Science and Technology of China, Hefei, Anhui 230026, China}

\author{X.~Dai,$^1$, K.~H.~Wu}
\affiliation{Beijing National Laboratory for Condensed Matter Physics, Institute of Physics, Chinese Academy of Sciences, Beijing 100190, China}

\author{Y.~Q.~Li}
\email[]{yqli@iphy.ac.cn}
\affiliation{Beijing National Laboratory for Condensed Matter Physics, Institute of Physics, Chinese Academy of Sciences, Beijing 100190, China}


\date{\today}

\begin{abstract}

We report that the finite thickness of three-dimensional topological insulator (TI) thin films produces an observable magnetoresistance (MR) in phase coherent transport in parallel magnetic fields. The MR data of  Bi$_2$Se$_3$ and (Bi$_{1-x}$Sb$_x$)$_2$Te$_3$ thin films are compared with existing theoretical models of parallel field magnetotransport. We conclude that the TI thin films bring  parallel field transport into a unique regime in which the coupling of surface states to bulk and to opposite surfaces is indispensable for understanding the observed MR.
The $\beta$ parameter extracted from parallel field MR can in principle provide a figure of merit for searching TI compounds with more insulating bulk than existing materials.

\end{abstract}



\maketitle

Much of the experimental effort on 3-dimensional (3D) topological insulators, an exotic class of quantum matter,~\cite{Hasan10,Qi11} has been devoted to obtaining samples with large bulk resistivities.~\cite{TItransport} This is essential to observing intrinsic transport properties of surface states~\cite{Ostrovsky10} and also a plethora  of fascinating effects emerging from hybrid structures combining TIs with other materials.~\cite{Fu08,Qi09,Fu09,Akhmerov09,Seradjeh09,Garate10,YuR11,Williams12} Observation of  surface transport was once very challenging, since the first identified TI materials~\cite{Fu07,Hsieh08,XiaY09,ChenY09,ZhangH09} tend to exhibit a conducting bulk.~\cite{Koehler73,Hor09,Butch10,Analytis10a} Recent breakthroughs in material design have produced TI compounds with bulk resistivities over 1\,$\Omega\cdot\mathrm{cm}$ (e.g.\ Bi$_2$Te$_2$Se and (Bi$_{1-x}$Sb$_x$)$_2$(Se$_{1-y}$Te$_y$)$_3$).~\cite{RenZ10,XiongJ12,RenZ11,Taskin11} Shubnikov-de Haas (SdH) oscillations observed in these materials are often,~\cite{Taskin09,Analytis10b,QuD10,Taskin11,Sacepe11,Petrushevsky12,RenZ12} if not always,~\cite{Butch10,Analytis10a,Bansal12} attributed to surface Dirac fermions. Another proven way  of observing  surface transport  in TI thin films is the use of electrical gating to deplete bulk carriers.  Large effects of gating on the longitudinal resistivity, the Hall effect, weak antilocalization, and SdH oscillations have been reported.~\cite{ChenJ10,Checkelsky11,ChenJ11,XiuF11,KongD11,Steinberg11,YuanHT11,KimD12,HeXY12} Although these advances facilitate observations of  surface transport, they also increase the difficulty in measuring residual bulk conductivity and evaluating its impact on surface transport, especially in thin films. A poorly conducting bulk  can still couple surfaces with opposite chiralities, which may damage fragile quantum states that rely on  topological protection. Deviation from exact quantization of anomalous Hall resistance in magnetically doped (Bi$_{1-x}$Sb$_x$)$_2$Te$_3$ thin films can also be attributed to bulk conductivity, which connects edge channels with opposite current directions and hence causes dissipation.~\cite{ChangCZ13} Therefore development of sensitive techniques for measuring bulk conductivity and related couplings is important  for making use of the interesting properties of the topological insulators.

Here we show that electron transport in magnetic fields applied parallel to TI thin films can serve as a convenient and sensitive probe for detecting bulk conductivity and couplings between surfaces and bulk or among surfaces.  It is based on manifestation of weak antilocalization (WAL) in parallel fields. At zero magnetic field  spin-momentum locking of surface Dirac fermions causes a $\pi$ Berry phase~\cite{Hsieh09} and suppression of backscattering.~\cite{Roushan09,Alpichshev10,ZhangT09} Its consequence in electron transport is referred to as the WAL.~\cite{Bergmann84,ChenJ10} In an ideal 3D TI thin film each of the top and bottom surfaces can be treated as a 2D system.  The principal effect of  a parallel field is to shift  the surface states in  momentum space, and this effect has no impact on transport in low magnetic fields.~\cite{Zyuzin11,note_on_edge,note_on_penetration} However Fig.\,1a shows that inter-surface and surface-bulk scatterings provide a finite thickness to electrons in a non-ideal TI because  electronic trajectories are no longer confined to the surface.~\cite{Tkachov11,ZhangW10,Mathur01}  The magnetic flux generated by the parallel field produces an Aharonov-Bohm phase which affects electrons when they make a closed loop~\cite{Altshuler81} and causes positive magnetoresistance (MR), similarly to the positive MR produced by a  perpendicular field.~\cite{Bergmann84} The latter is usually much more pronounced, and it has been a subject of intensive studies.~\cite{ChenJ10,Checkelsky11,ChenJ11,Steinberg11,HeHT11,KimY11,KimD12b,Tkachov11,LuHZ11,WangJ11,Garate12,LeeJ12}  In contrast, the parallel field MR has not received systematic experimental effort despite a few theoretical predictions on novel quantum effects induced by the parallel magnetic fields.~\cite{Tkachov11,Zyuzin11,Pershoguba12}

The weak (anti)localization effect  in parallel field transport was first predicted in 1981 by Altshuler and Aronov (AA) for dirty metal films~\cite{Altshuler81}, and  was later generalized to cleaner films~\cite{Dugaev84,Beenakker88} and bilayer systems.~\cite{Raichev00} In this paper we show that none of these models developed for topologically trivial systems can fully describe the parallel field MR observed in Bi$_2$Se$_3$ and (Bi$_{1-x}$Sb$_x$)$_2$Te$_3$ thin films. Our data point to \textit{qualitatively} different regimes for  phase coherent transport in parallel magnetic fields, and suggest that the essential ingredients include the existence of  surface states, their coupling to the bulk, and coupling between opposite surfaces. Knowledge obtained from perpendicular field transport alone is insufficient to provide a satisfactory description of the complicated electron systems encountered in this work. For instance, the widely used two-channel model~\cite{ChenJ11,Garate12,Bansal12,KimD12b} fails to explain the ambipolar transport in (Bi$_{1-x}$Sb$_x$)$_2$Te$_3$ thin films, and the parallel field transport suggests the importance of bulk conductivity in these films.

\begin{figure}
\includegraphics*[width=8.5 cm]{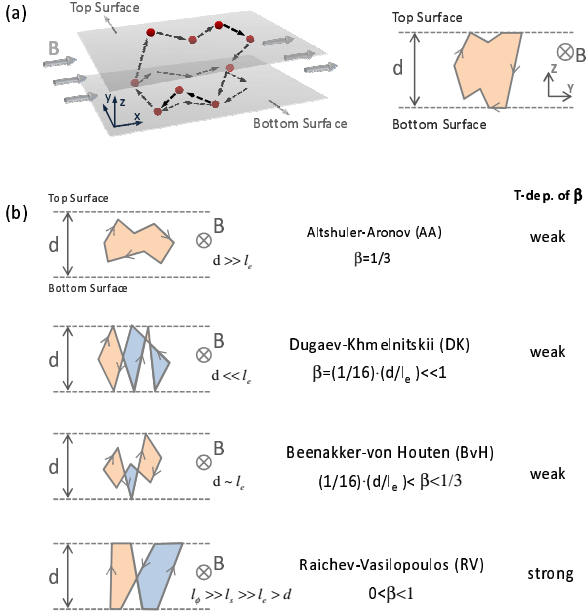}
\caption{\label{Fig1} (color online) (a) A possible closed-loop electron trajectory in a topological insulator thin film (left) and its  cross-section projected perpendicular to the parallel magnetic field (right). The electron's motion may include transport on the surface and in the bulk, as well as surface-bulk and the inter-surface scatterings. (b) Cross-section view of  electron trajectories illustrating the Aharonov-Bohm (AB) phase induced by a parallel magnetic field in various transport regimes. Shown from top to bottom are AA~\cite{Altshuler81}, DK~\cite{Dugaev84}, BvH~\cite{Beenakker88}, and RV~\cite{Raichev00} regimes. See Eq.\,(\ref{eq:parallel}) for a definition of $\beta$.}
\end{figure}


Fig.\,1b illustrates several regimes of  parallel field transport. These include the AA regime for thin films with mean free path much smaller than the film thickness ($l_e \ll d$),~\cite{Altshuler81} the Dugaev-Khmelnitskii (DK) regime for clean metal films ($l_e \gg d$),~\cite{Dugaev84} and the Beenakker-van Houten (BvH) regime,~\cite{Beenakker88} which describes the crossover  between the AA and DK regimes. The correction to the conductivity in all three regimes can be written in a unified form:
\begin{equation}
\label{eq:parallel}
\Delta\sigma_{||}(B)\simeq
\alpha\frac{-e^2}{2\pi^2\hbar}
\ln\left(1+\beta\frac{e d^2 }{4\hbar B_\phi}B^2\right),
\end{equation}
where $\Delta\sigma_{||}(B)\equiv\sigma_\mathrm{xx}(B_{||})-\sigma_\mathrm{xx}(0)\simeq -\mathrm{MR}/\rho_\mathrm{xx}(0)$.
The parameter $\alpha$ takes values of $1/2$ and $-1$ respectively for weak antilocalization and weak localization.
In these traditional single layer systems the upper bound of $\beta$ is reached in the AA regime with $\beta_\mathrm{AA}=1/3$ while the DK regime gives the lower limit\ $\beta_\mathrm{DK}=(1/16)d/l_e \ll 1$.~\cite{Supplement}
The dephasing field  $B_\phi$ can be obtained by fitting the perpendicular field magnetoconductivity (MC) to the Hikami-Larkin-Nagaoka (HLN) equation,~\cite{Hikami80} which in both strong  and weak limits of the spin-orbit coupling (SOC)  simplifies to:
\begin{equation}
\label{eq:HLN}
\Delta\sigma_\perp(B)\simeq
\alpha\frac{-e^2}{2\pi^2\hbar}
\left[
\psi\left(\frac{1}{2}+\frac{B_\phi}{B}\right)
-\ln\left(\frac{B_\phi}{B}\right)
\right],
\end{equation}
where $\psi(x)$ is the digamma function.

A similar magnetoconductivity exists also in coupled bilayer systems, but its magnitude depends on the conductivity and dephasing time of each layer ($\sigma_\mathrm{xx,i}$ and $\tau_{\phi,i}$, $i$=1,2), as well as the interlayer tunneling (or transition) time $\tau_s$. For symmetric bilayers (with $\sigma_\mathrm{xx,1}$=$\sigma_\mathrm{xx,2}$ and $\tau_{\phi,1}$=$\tau_{\phi,2}$=$\tau_\phi$) $\Delta\sigma_{||}(B)$ follows the same form as Eq.\,(1) with $\beta=2(1+s)/(1+2s)-\ln(1+2s)/s$ and $s=\tau_\phi/\tau_s$, according to Raichev and Vasilopoulos (RV).~\cite{Raichev00} In the weak coupling limit ($\tau_\phi \ll \tau_s$) $\beta$ is suppressed quadratically $\beta\simeq(4/3)(\tau_\phi/\tau_s)^2\ll 1$, whereas in the strong coupling limit ($\tau_\phi\gg\tau_s$) $\beta$ approaches  1 (see supplemental information~\cite{Supplement}).


Fig.\,2 shows the magnetotransport results of Bi$_2$Se$_3$ thin films with thicknesses of 7-45\,nm for both perpendicular and parallel field orientations. The parallel field MC has a \textit{pronounced} dependence on thickness, and its field dependence is drastically different from that of the perpendicular MC (Fig.\,2a).  Hall measurements show that the sheet carrier density $n_s$ is in the range 1.8-4.6$\times10^{13}$\,cm$^{-2}$,~\cite{Supplement} which indicates that the Fermi level is located in the bulk conduction band, in agreement with previous studies.~\cite{XiaY09,Analytis10a,KimY11} The perpendicular field MC can be fitted to the HLN equation(Fig.\,2b), and $\alpha$ is  close to 1/2 for all films (Fig.\,2c). This is  consistent with previous measurements of Bi$_2$Se$_3$ across a large range of electron densities,~\cite{ChenJ11} mobilities,~\cite{ChenJ11} and film thicknesses.~\cite{KimY11} At the electron densities encountered in this work both  surface and bulk carriers are expected to contribute substantially to transport. Nonetheless the WAL manifested in perpendicular field transport does not show any qualitative difference from topologically trivial thin films with strong SOC,~\cite{Bergmann84} such as  Au thin films.~\cite{Hoffmann82}  It has been suggested that strong scattering between surface and bulk states makes the system behave as a single channel system  when the dephasing rate is much smaller than the surface-bulk scattering rate.~\cite{ChenJ11,Steinberg11,Garate12,Supplement} An angle-resolved photoemission spectroscopy (ARPES) experiment has found evidence for strong surface-bulk scattering in Bi$_2$Se$_3$ when the Fermi level is located in the bulk conduction band.~\cite{Park10}

\begin{figure}
\includegraphics*[width=8.5 cm]{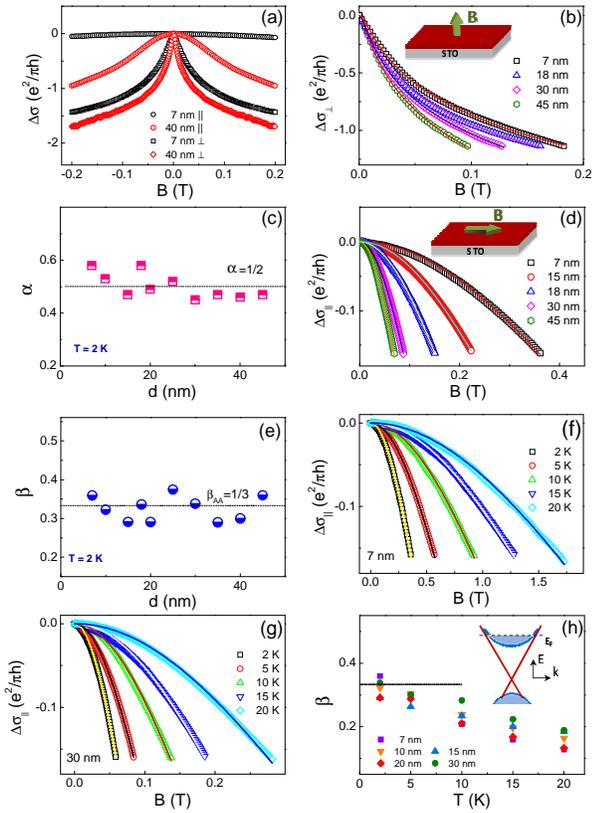}
\caption{\label{Fig2} (color online) Magnetotransport data of Bi$_2$Se$_3$ thin films in perpendicular and parallel magnetic fields. (a) Magnetoconductivity (MC) in perpendicular fields $\Delta\sigma_{\perp}(B)$, and MC in parallel fields $\Delta\sigma_{\parallel}(B)$ for films with thickness $d=$ 7 and 40\,nm. (b) $\Delta\sigma_{\perp}(B)$ for various thicknesses (symbols: experiment, lines: fits to the HLN equation). (c) Extracted parameter $\alpha$ vs. the film thickness $d$. (d) $\Delta\sigma_{\parallel}(B)$ for various thicknesses [symbols: experiment, lines: fits to Eq.\,(1)]. (e) Thickness dependence of $\beta$. The data in panels (a-e) were obtained at $T\approx 2$\,K. (f-g) $\Delta\sigma_{\parallel}(B)$ at various temperatures for films with (g) $d=$7\,nm and (h) $d=$30\,nm. (h) $T$-dependence of $\beta$ for $d$=7-30\,nm. The inset shows a schematic band diagram of the Bi$_2$Se$_3$ films.}
\end{figure}

Fig.\,2d illustrates the strong thickness dependence of the parallel field MC. The data are fitted to Eq.\,(1) with $\alpha$ fixed at 1/2. Since $B_\phi$ can be obtained from the HLN fits, $\beta$ is the only free parameter.
In Fig.\,2e we fix  $T\approx 2$\,K and  show that for all thicknesses studied here $\beta$ is close to 1/3, the value predicted by AA for topologically trivial dirty metal films. The fact that $\beta$ is nearly constant implies that $\Delta\sigma_\parallel(B)$ is proportional to $d^2$ in low fields [see Eq.\,(\ref{eq:parallel})].

Even though these observations  seem consistent with the AA prediction for a single bulk layer, this is only a coincidence, because the AA regime requires the surfaces be separated by many scattering lengths, namely $d \gg l_e$.   In each of the Bi$_2$Se$_3$ thin films, however, $l_e$ for the bulk carriers is found to be comparable to the film thickness $d$ (see supplemental information~\cite{Supplement}).  The $l_e/d$ ratio lies in the  BvH regime where theory predicts that $\beta$ is considerably smaller  than the measured value  $\beta\approx1/3$.
Furthermore theories of single layer systems predict that  $\beta$ depends only on the ratio $l_e/d$ and should be nearly independent of $T$ at low temperatures because $l_e$ is mainly determined by  impurity scattering. In contrast, Fig.\,2h shows that $\beta$ decreases by a factor of about two as $T$ is increased from 2\,K to 20\,K. This further proves that the parallel field MC in Bi$_2$Se$_3$ cannot be explained by traditional transport models developed for thin films.

The existence of the surface states and their couplings to bulk may be key to the observed parallel field MC.  A sizable portion of the  current is carried by the surface states in these films, and this inhomogenous current distribution can account for the discrepancy between the BvH prediction and the much larger experimental values.~\cite{Beenakker88} The $T$-dependence of $\beta$ can be attributed to the variation in the ratio $\tau_\phi/\tau_\mathrm{sb}$, where $\tau_\mathrm{sb}^{-1}$ is the scattering rate between surface and bulk carriers and $\tau_\phi$ is the  dephasing time. As suggested in the ARPES experiment,~\cite{Park10} $\tau_\mathrm{sb}^{-1}$ is dominated by sample disorder at low $T$, and hence is expected to be nearly $T$-independent. Electron-electron interactions are the leading source of electron dephasing at low $T$ and hence $\tau_\phi^{-1}\propto T$.~\cite{Altshuler85,LinJJ02}  This implies that increasing $T$ causes the surface states to decouple from the bulk, resulting in decrease in the finite thickness effect.

\begin{figure}
\includegraphics*[width=8.5 cm]{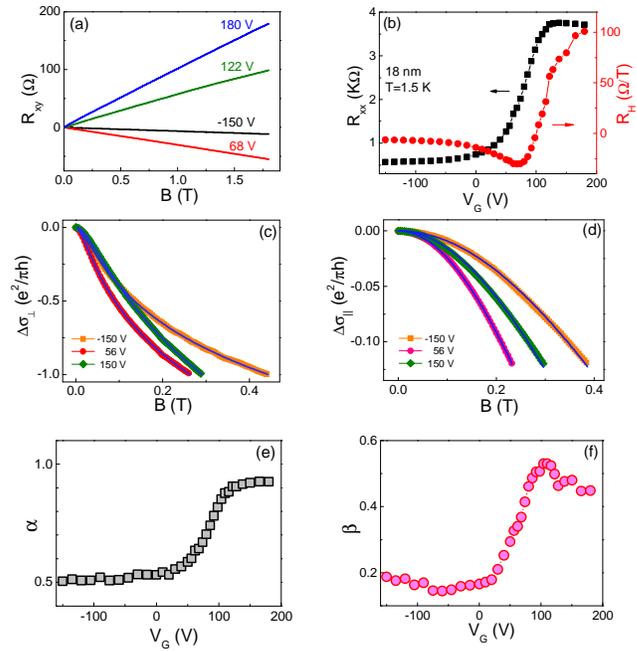}
\caption{\label{Fig3} (color online) Magnetoconductivity of an 18\,nm thick (Bi,Sb)$_2$Te$_3$ film at T=1.5\,K. (a) Hall resistance $R_\mathrm{xy}$ at various gate voltages. (b) Gate voltage dependencies of the longitudinal resistance  $R_\mathrm{xx}$ (squares) and the Hall coefficient $R_H$ (circles). In (a-b), The sign of $R_\mathrm{xy}$ and $R_H$ is set to $+1$ for electrons.  (c) MC in perpendicular fields for various gate voltages. (d) MC in parallel fields. (e) Gate voltage dependence of $\alpha$. (f) Gate voltage dependence of $\beta$.}
\end{figure}

In Bi$_2$Se$_3$ both the high density of selenium vacancies and various surface doping effects~\cite{Hor09,Analytis10a} makes it difficult for back-gating to effectively suppress bulk transport.
This difficulty is remedied in (Bi$_{1-x}$Sb$_x$)$_2$Te$_3$ compounds,~\cite{KongD11,ZhangJ11} which display much improved gate tunability.~\cite{KongD11,HeXY12}
Ambipolar transport can be obtained reliably in (Bi,Sb)$_2$Te$_3$ films grown epitaxially on SrTiO$_3$. Fig.\,3 shows transport data for an 18\,nm (Bi,Sb)$_2$Te$_3$ sample at $T=1.5$\,K. At $V_G=0$ the sample is p-type, but at large positive gate voltages ($V_G>80$\,V) it enters the ambipolar regime where both electrons and holes participate in  transport.~\cite{ChenJ11,Steinberg11,Checkelsky11,KimD12,HeXY12}  The  hallmarks of this regime include the appearance of extrema of $R_H$,  reversal of $R_H$'s sign, a non-linear Hall resistance, and a maximum in $R_\mathrm{xx}$.~\cite{HeXY12} These features can be explained on a qualitative level by a model with two channels: an electron layer near the bottom interface and a hole layer near the top surface, separated by an at least partially depleted bulk layer.  This resembles a coupled bilayer system.~\cite{Raichev00,Garate12}  If the depletion layer is  good enough, i.e.\ if  the inter-layer tunneling or transition rate $\tau_s^{-1}$ is comparable to or smaller than the dephasing rate $\tau_\phi^{-1}$,  $\alpha$ will shift from $1/2$ to larger values $1/2 < \alpha \leq 1$.~\cite{Garate12} This is indeed observed in the (Bi$_{1-x}$Sb$_x$)$_2$Te$_3$ sample shown in Fig.\,3 as well as  in gated Bi$_2$Se$_3$ samples.~\cite{ChenJ11,Steinberg11,KimD12b}

\begin{figure}
\includegraphics*[width=8.5 cm]{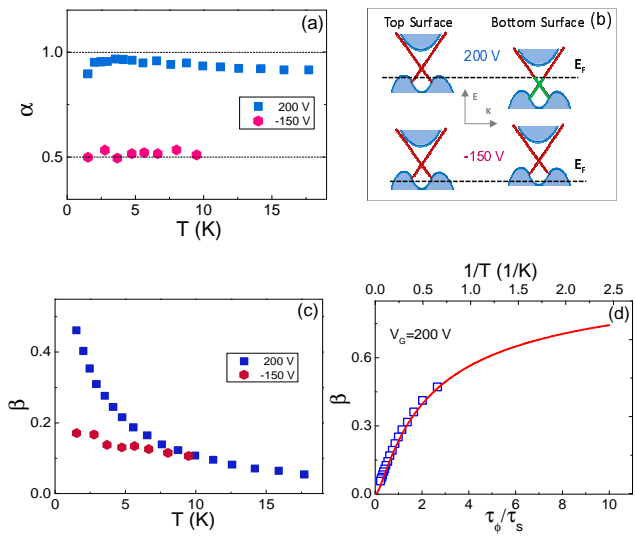}
\caption{\label{Fig4} (color online) (a) Temperature dependence of $\alpha$ at $V_G=-150$\,V (hexagons) and $+200$\,V (squares). (b) Schematic band diagrams for these two gate voltages. (c) $T$-dependencies of $\beta$ at $V_G=-150$\,V and  $+200$\,V. (d) $\beta$  plotted as a function of $1/T$ (squares, $V_G=+200$\,V) and as a function of the coupling strength $\tau_\phi/\tau_s$ (line, symmetric bilayer in RV theory~\cite{Raichev00}).}
\end{figure}

 The temperature dependence shown in Fig.\,4 may shed more light on the nature of the magnetotransport. We report data in two different transport regimes: first with the gate voltage fixed at $-150$\,V  placing the Fermi energy in the bulk valence band, and second at $+200$\,V producing ambipolar transport. When the Fermi energy is in the bulk valence band the bulk $\alpha$ is close to $1/2$ and $\beta$ has a weak temperature dependence, similarly to Bi$_2$Se$_3$ thin films with $E_F$ located in the bulk conduction band. This is in stark contrast to the strong temperature dependence seen in the ambipolar regime, where $\beta$ decreases by a factor of about eight as $T$ is increased from 1.5\,K to 20\,K.
 As mentioned above, such a strong $T$-dependence of $\beta$ is not possible in single layer systems (whether in AA, DK, or BvH regimes) but is possible in a bilayer system.  In the RV model of a bilayer system $\beta$  increases monotonically from 0 to 1 as the inter-channel coupling strength $s=\tau_\phi/\tau_s$ increases. Assuming $\tau_s$ is $T$-independent, we would expect $s\propto 1/T$ at low $T$.

The two-channel model explains at least qualitatively the individual features of transport in (Bi,Sb)$_2$Te$_3$, but nonetheless fails to explain transport in the perpendicular and parallel fields \textit{simultaneously}.  In this 18\,nm sample $\alpha$ exceeds 0.9 when $V_G>120$\,V, which in the two-channel RV model implies that the coupling between the top and bottom layers is rather weak: \ $s=\tau_\phi/\tau_s \leq 0.6$. Such small coupling would lead to $\beta \leq 0.1$.  In contrast, Fig.\,\ref{Fig4}d shows that $\beta>0.4$ extracted from experiment is corresponding to rather large coupling strengths in the two-channel RV model.

The magnetotransport data shown above suggest that transport in the ambipolar regime of the (Bi,Sb)$_2$Te$_3$ sample is beyond the description of the coupled two-channel model. The large $\beta$ parameters can only be attributed to an extra source of MC in addition to the two channels on (or near) the top and bottom surfaces. The residue bulk conductivity is very likely responsible for the third channel, which leads to extra parallel field MC and underestimation of the coupling strength $s$ from the perpendicular field data. The bulk states also provide venues for couplings between the top and bottom surfaces in the 18\,nm thick sample, in which the thickness is too large for significant inter-surface tunneling.~\cite{ZhangW10} The sizable bulk conductivity inferred from this work is consistent with recent measurements of quantum anomalous Hall resistance $R_\mathrm{AH}$ in Cr doped (Bi,Sb)$_2$Te$_3$ thin films,~\cite{ChangCZ13} in which a bulk resistance on the order of 10\,k$\Omega$/square at $T=1.5$\,K can be estimated from the deviation of $R_\mathrm{AH}$ from $R_K\equiv h/e^2$. Searching for TIs with more insulating bulk will certainly help to achieve more precise quantization of $R_\mathrm{AH}$. A key signature for ideal TI thin films would be observing $\beta$ values approaching to zero.

In summary, we have shown that  parallel field transport can provide unique insight into electron transport in 3D TIs that cannot be obtained from perpendicular field transport alone. In particular, the $\beta$ parameter  extracted from the parallel field MR can in principle allow measurement of   surface-bulk  scattering and of the bulk conductivity.~\cite{Vincent13}
The $\beta$ parameter is expected to be a figure of merit for characterization of 3D TI materials. Further theoretical work will be valuable to establish this parallel field method as a quantitative tool for study of TIs.

We are grateful to D. Culcer, D. Goldharber-Gordon, K. He, A. D. Mirlin, P. Ostrovsky, and P. Xiong for helpful discussions.  We acknowledge valuable support from the Laboratory of Microfabrication at the IOP, and financial support from National Basic Research Program of China (Project Nos.\ 2012CB921703 and 2009CB929101), National Science Foundation of China (Project No.\ 91121003), Chinese Academy of Sciences.


\begin{thebibliography}{}
\bibitem{note_on_edge} In this work we neglect the magnetoconductivity of the side surfaces since the Hall bars' width of 50\,$\mu$m is three orders of magnitude larger than their thickness ($\leq 50$\,nm).
\bibitem{note_on_penetration}  Ref.\protect{~\cite{Tkachov11}} finds that the finite penetration depth $\lambda$ of the surface state wavefunction leads to a positive parallel field MR proportional to $\ln(1+b B^2)$ with $b\propto\lambda^2$. However, this effect is much smaller than the effects discussed in the main text as well as surface roughness effect,~\protect{\cite{Mathur01}} since $\lambda$ is only about 2\,nm or less for typical TIs.~\protect{\cite{ZhangW10}}
\end{thebibliography}
\end{document}